\documentclass[twocolumn,10pt]{article}

\usepackage[utf8]{inputenc}
\usepackage{graphicx}
\usepackage{amsmath}
\usepackage{amssymb}
\usepackage{amsfonts}
\usepackage{booktabs}
\usepackage{tabularx}
\usepackage{siunitx}
\usepackage{xcolor}
\usepackage{lineno}
\usepackage{verbatim}
\usepackage[colorlinks=true,linkcolor=blue,citecolor=blue]{hyperref}
\usepackage{doi}
\usepackage[english]{babel}
\usepackage{microtype}
\usepackage{floatrow}
\hypersetup{breaklinks=true}
\usepackage{xurl}
\usepackage{lipsum}
\usepackage[letterpaper,top=1.8cm,bottom=2.0cm,left=1.6cm,right=1.6cm]{geometry}
\usepackage{newtxtext,newtxmath} 
\usepackage[numbers,square,sort&compress]{natbib}
\usepackage{authblk}

\definecolor{myblue}{HTML}{1f77b4}
\definecolor{mygreen}{HTML}{2ca02c}
\definecolor{myred}{HTML}{d62728}
\definecolor{myorange}{HTML}{ff7f0e}

\newcommand{\sep}{; }

\title{New muon transfer cell for muonic x-ray spectroscopy on microgram-quantity targets based on liquid hydrogen}

\author[1,3]{M.~Enciu\thanks{Corresponding author.}}
\author[1,2,3]{A.~Antognini}
\author[1,2,3]{K.~Kirch}
\author[2,3]{A.~Knecht}
\author[1]{K.~Lehrke}
\author[1]{J.~V\"olkert}
\author[1]{Y.~Zhuang}

\affil[1]{Institute of Particle Physics and Astrophysics, ETH Z\"urich, 8093 Z\"urich, Switzerland}
\affil[2]{PSI Center for Neutron and Muon Sciences, 5232 Villigen PSI, Switzerland}
\affil[3]{National Centre of Competence in Research Muoniverse, 5232 Villigen PSI, Switzerland}

\begin{document}

\twocolumn[
\maketitle

\begin{abstract}
The muon transfer method allows access to radioactive/rare atomic nuclei for determining their charge radii via muonic x-ray spectroscopy. In this study we present a new muon transfer cell based on liquid hydrogen optimized for low-energy x-ray spectroscopy by employing a thin-walled Mylar cell for the container of the liquid hydrogen. Based on simulations, the present liquid hydrogen cell achieves competitive efficiencies compared to existing devices using the muon transfer method. Additionally, the liquid-hydrogen-based muon transfer setup provides a short timescale for the diffusion process of the muonic hydrogen atoms in the cell, well below 200~ns.
\end{abstract}

\vspace*{\baselineskip}
\vspace*{\baselineskip}
\vspace*{\baselineskip}
]

\section{Introduction}
Nuclear charge radii can be measured by different means such as elastic scattering with leptonic probes or spectroscopy of atomic transitions in electronic and muonic atoms~\cite{Nrtershuser2022}. The implications of precision charge radii determination range from benchmarking state-of-the-art ab initio nuclear theory models~\cite{Hergert2020,Arthuis2026}, learning about the link of nuclear radii to the equation of state of the nuclear matter~\cite{Lattimer2021,Novario2023}, and even testing deviations from the standard model of particle physics~\cite{Safronova2018,Okumura2023}. X-ray spectroscopy of muonic atoms, i.e. atoms with electrons replaced by a muon, is a well-established method for extracting absolute charge radii with high precision and has been applied to almost all stable nuclei~\cite{Knecht2020}.

For studying unstable nuclei, however, traditional muonic x-ray spectroscopy methods involving a muon beam bombarding a thick target cannot be applied. A muon transfer cell based on hydrogen gas was developed by the muX collaboration~\cite{Wauters2021} in order to bring muonic x-ray spectroscopy measurements on microgram target materials within reach~\cite{Adamczak2023}. The muon transfer method is based on the pioneering work in Refs.~\cite{Strasser1999,Strasser2001}.\\
In the muon transfer method, the muon is first stopped in a hydrogen–deuterium mixture and captured on a hydrogen atom. Upon encountering deuterium, the muon transfers to form muonic deuterium ($\mu d$) with eV-range kinetic energy, whose diffusion is strongly enhanced by the Ramsauer–Townsend effect~\cite{Adamczak2023}. The neutral $\mu d$ atom then diffuses to the target of interest, transferring the muon to a higher-$Z$ atom via the chain $\mu^- \rightarrow \mu p \rightarrow \mu d \rightarrow \mu Z$.\\
While the muX collaboration~\cite{Wauters2021} focuses on heavy nuclei studies, the QUARTET collaboration~\cite{Ohayon2024} aims at improving the precision of charge radii for the low-medium-Z elements ($Z\leq12$) in combination with the use of metallic magnetic calorimeters (MMC) reaching a resolving power of up to 6000~\cite{Unger2024}.

In this technical design report we present the development of a new device for muonic x-ray spectroscopy on microgram target materials via the transfer method operating with liquid hydrogen (LH$_2$). In Section 2 we discuss in details the processes happening in the muon transfer cell, we present Monte Carlo simulations performed to evaluate the muon transfer rate for different configurations of geometry and hydrogen properties, and we conclude by comparing the transfer cells based on liquid hydrogen and gaseous hydrogen. In Section 3 we describe the mechanical design of the cryogenic LH$_2$ system, followed by Section 4, where we detail the integration of the cryogenic system with the full experimental setup including two configurations: MMC detectors and a high-purity Ge (HPGe) detector array. In Section 5, we present the commissioning and the performance of the LH$_2$ target system, and lastly we discuss the conclusions and the physics outlook with the newly developed LH$_2$ transfer cell.

\section{Muon transfer in liquid hydrogen}
\subsection{The muon transfer process}
The main processes happening inside the muon transfer cell are the following: the muons ($\mu^-$) enter the cell and lose energy down to a few eV; the muons get captured on the hydrogen isotopes; muonic hydrogen atoms are transported towards the microgram target by diffusion; the diffusion process is enhanced by a small deuterium addition into hydrogen; the muon is transferred from the muonic hydrogen isotopes to higher-$Z$ isotope in the target of interest as follows $\mu \rightarrow \mu p \rightarrow \mu d \rightarrow \mu Z$; the muons cascade down the muonic energy levels within the target atoms emitting Auger electrons and muonic x rays; finally, the emitted characteristic x-rays from the target of interest are detected by a high-resolution calorimeter. The process is detailed in~\cite{Adamczak2023}.

\begin{figure}[t]
    \centering
    \includegraphics[width=1\linewidth]{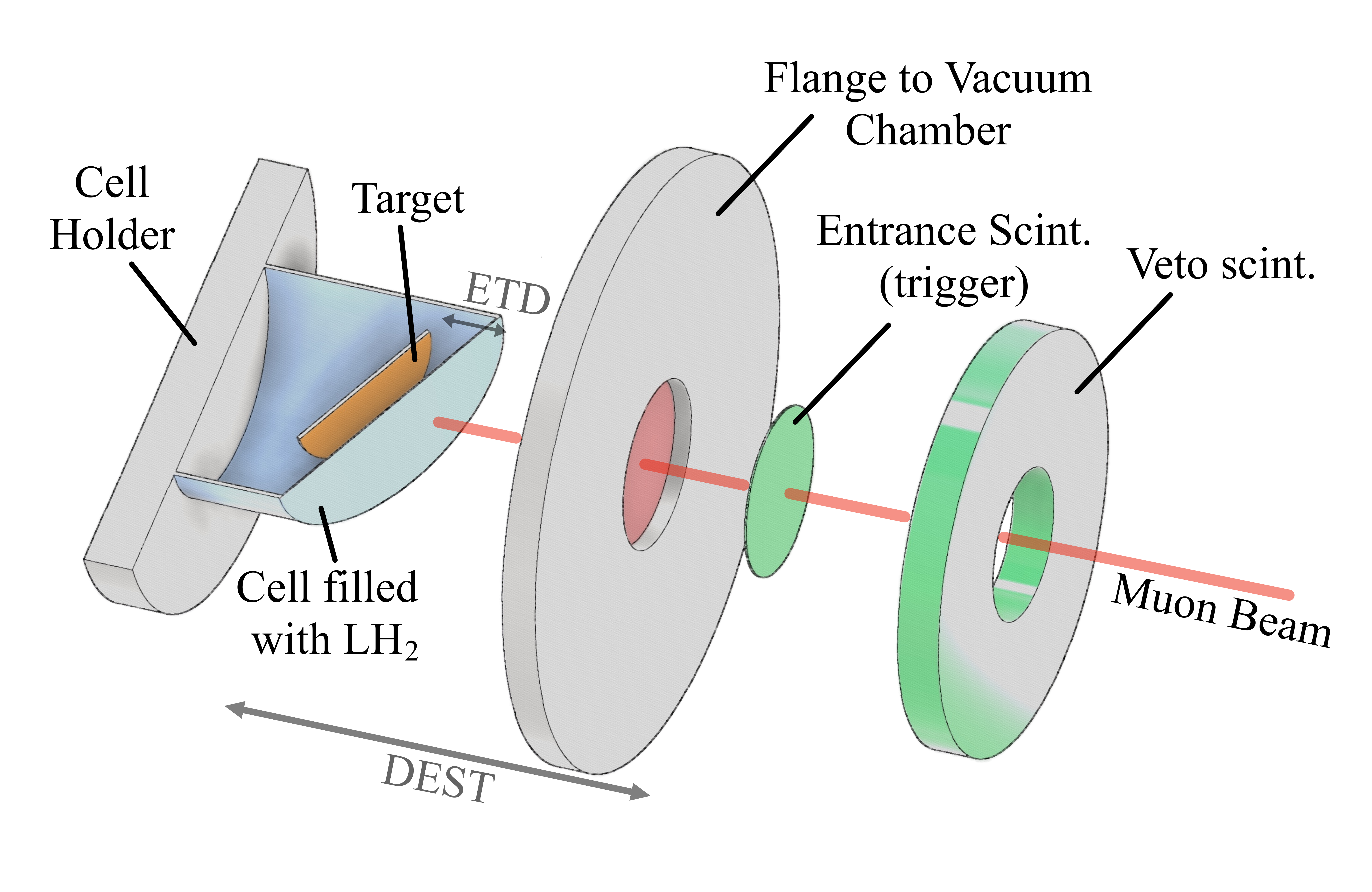}
    \caption{Schematic drawing of the geometry used for the muon transfer simulation. Two selected geometrical parameters, $ETD$ and $DEST$, are marked and discussed in the text.}
    \label{fig:sim-geometry}
\end{figure}

\subsection{Muon transfer simulations and results}
The performance of the muon transfer process in liquid hydrogen was evaluated based on the simulation code developed in Ref.~\cite{Adamczak2023}. The simulation includes the following processes: muons losing energy until they stop; if stopped in the hydrogen medium, the muon gets captured, forming muonic hydrogen, directly de-exciting into its atomic ground state; the diffusion of muonic hydrogen is simulated in the hydrogen medium containing the admixture of deuterium; encountering a deuterium atom, the muon transfer from hydrogen to deuterium is favored with the excess binding energy resulting in a boost in kinetic energy; the diffusion of muonic deuterium is further simulated until it reaches the target; the track is killed when it reaches the target and gets counted for the muon transfer efficiency; tracks which follow other reaction processes than described here or exceed the muon lifetime are killed too, without being counted for efficiency. 

The muon momentum was treated as a variable parameter, but a momentum spread of 4\% (FWHM) and a Gaussian spatial distribution with $\sigma_{\text{x}}=\sigma_{\text{y}}=10~mm$ were kept constant through the simulation. In this simulation the deuterium concentration, $c_{\text{D}}$, was also treated as a variable parameter. 
\begin{figure}
    \centering
    \includegraphics[width=0.95\linewidth]{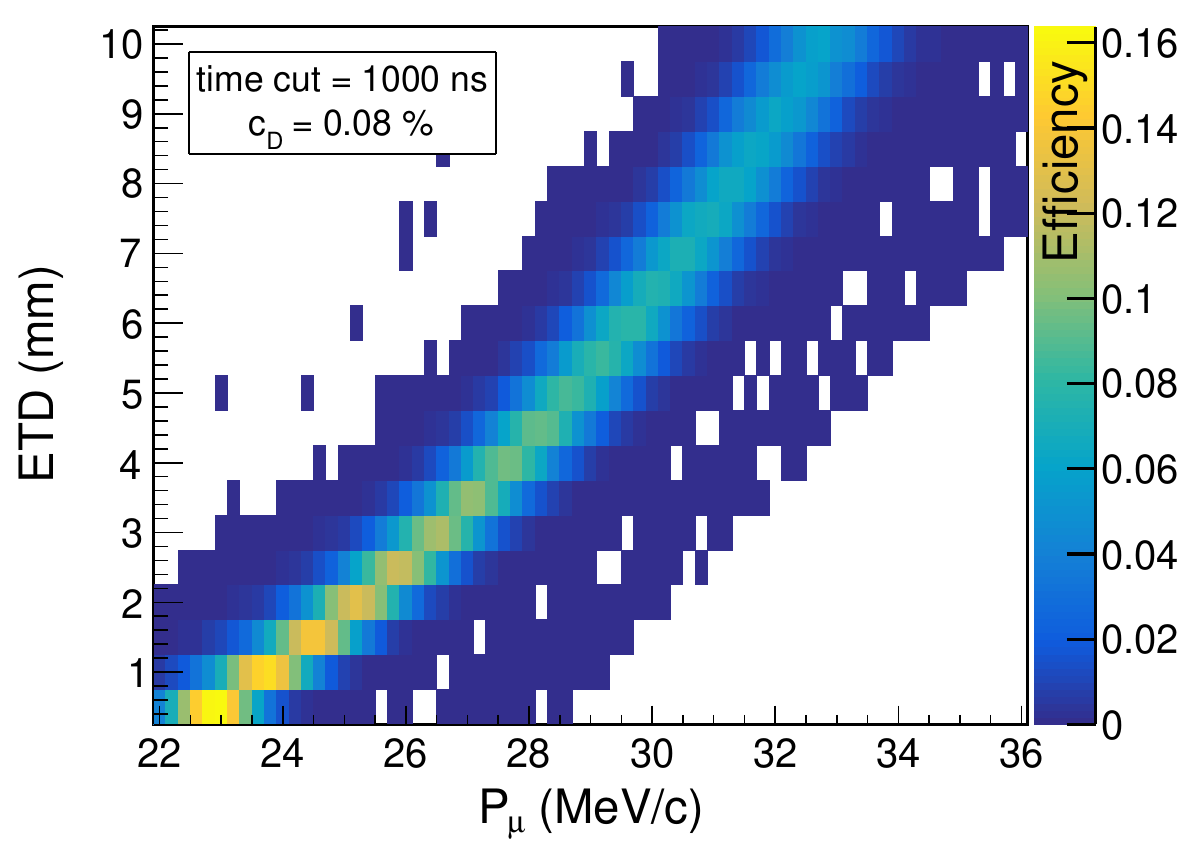}
    \caption{The relation between the muon momentum ($P_{\mu}$) and the ETD parameter. The z-axis shows the total efficiency. The  plot is shown for DEST = 25~mm, $c_{\text{D}}=0.08~\%$ and only muonic atoms of hydrogen reaching the target within 1000~ns are counted, see text for details.}
    \label{fig:EDT-muP}
\end{figure}
The geometry input for the simulation contains the following elements visualized in Figure~\ref{fig:sim-geometry}: (1) Veto scintillation detector (with an opening of 18~mm in diameter); (2) Entrance scintillation detector (20~mm diameter, distance from Veto = 25~mm); (3) Interface, 50~$\mu$m-thick Mylar foil, between the vacuum chamber enclosing the scintillation detectors and the vacuum chamber enclosing the liquid hydrogen target (20~mm diameter, distance from Entrance scint. = 17.5~mm, shown with an off-red color in Figure~\ref{fig:sim-geometry}); (4) The cell made out of 175~$\mu$m-thick thermoformed Mylar, filled with liquid hydrogen (dimensions detailed in Section~\ref{sec:Cell}); (5) The measured target (20-mm-diameter disk); (6) The stainless steel flange support of the cell (diameter of 57~mm and thickness of 8~mm).

\begin{figure*}
    \centering
    \includegraphics[width=0.912\linewidth]{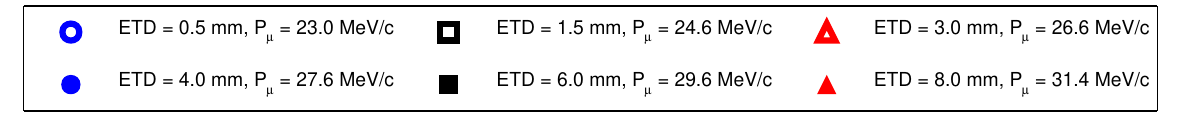}\\
    \includegraphics[width=0.475\linewidth]{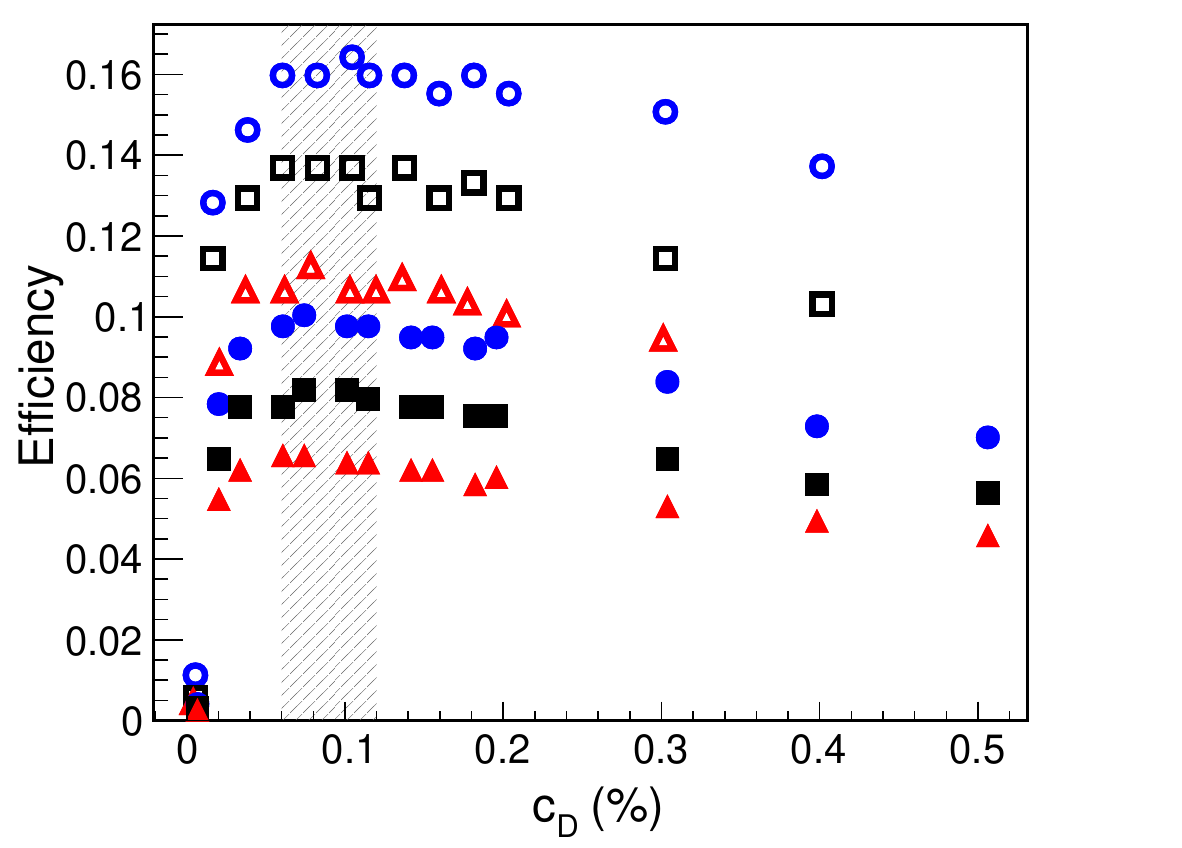}
    \hfill
    \includegraphics[width=0.475\linewidth]{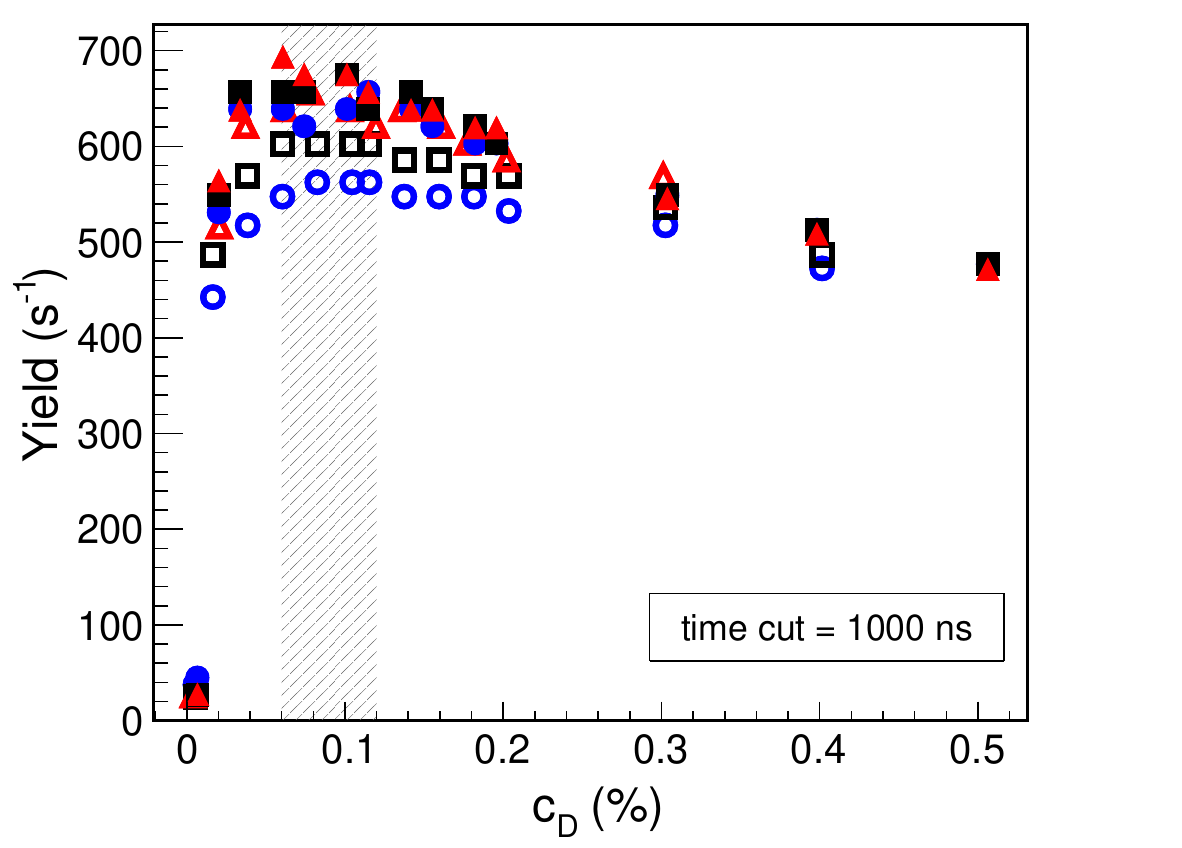}
    \caption{Efficiency (left) and yield (right) as function of the deuterium concentration ($c_{\text{D}}$) for several optimized pairs of (ETD, $P_{\mu}$). The hatched areas mark the interval $c_{\text{D}} =[0.06,0.12]~\%$ in which the maximum efficiency and yield are found.}
    \label{fig:eff-Yield_cD}
\end{figure*}
Several geometrical parameters were varied during this study to see their impact on the muon transfer efficiency. The target and the front face of the cell could be tilted and it is found that, for increased efficiency, the tilt angle of both of them needs to be the same, regardless of the angle value, to keep a constant thickness of LH$_2$ layer. The tilt angle value impacts the efficiency only due to minimizing the effective target surface area seen by the muon beam. The diameter of the cell and the diameter of the target must be similar to the muon beam size; if they are much smaller than that, the efficiency drops, but if they are much larger, the efficiency is not affected. The optimum liquid hydrogen layer thickness between the entrance into the cell and the target, visualized in Figure~\ref{fig:sim-geometry} and denoted as ETD, is linked to the range of muons in the dense fluid and hence dependent on the momentum of the incident muon beam as shown in Figure~\ref{fig:EDT-muP}. The total length of the cell is kept considerably longer than ETD in order to lower the possibility of muons which did not reach the target to reach other materials such as the stainless steel flange. The cell diameter is also chosen slightly larger than the target and the beam size to lower the probability of muons to reach the cell walls; instead, the muons will stop and diffuse through the hydrogen-deuterium mix until decay. 

The thickness and material of the foil mounted at the interface between the two vacuum chambers (`Flange to Vacuum Chamber' in Figure~\ref{fig:sim-geometry}), in this case 50-$\mu$m-thick Mylar foil, plays a role in straggling of the muon beam as it passes through it; the muons are randomly scattered away from the direction towards the target if a larger thickness is chosen. Finally, the distance between the entrance scintillator and the target cell, visualized in Figure~\ref{fig:sim-geometry} and denoted as DEST, has a similar effect: if chosen too large, the muon beam can diverge more and the entrance scintillator can trigger more events in which the muon does not reach the target, lowering the effective efficiency. The target tilt angle, foil thickness and DEST contribute to a purely geometrical efficiency, while the ETD, muon momentum and the deuterium concentration contribute to the muon transfer efficiency. For all results of the simulation we will keep DEST = 25~mm, the same as in the case of the hydrogen gas cell in Ref.~\cite{Adamczak2023}, for comparison, as larger DEST values would drop the total efficiency due to the geometric reasons only.
\begin{figure}
    \centering
    \includegraphics[width=0.95\linewidth]{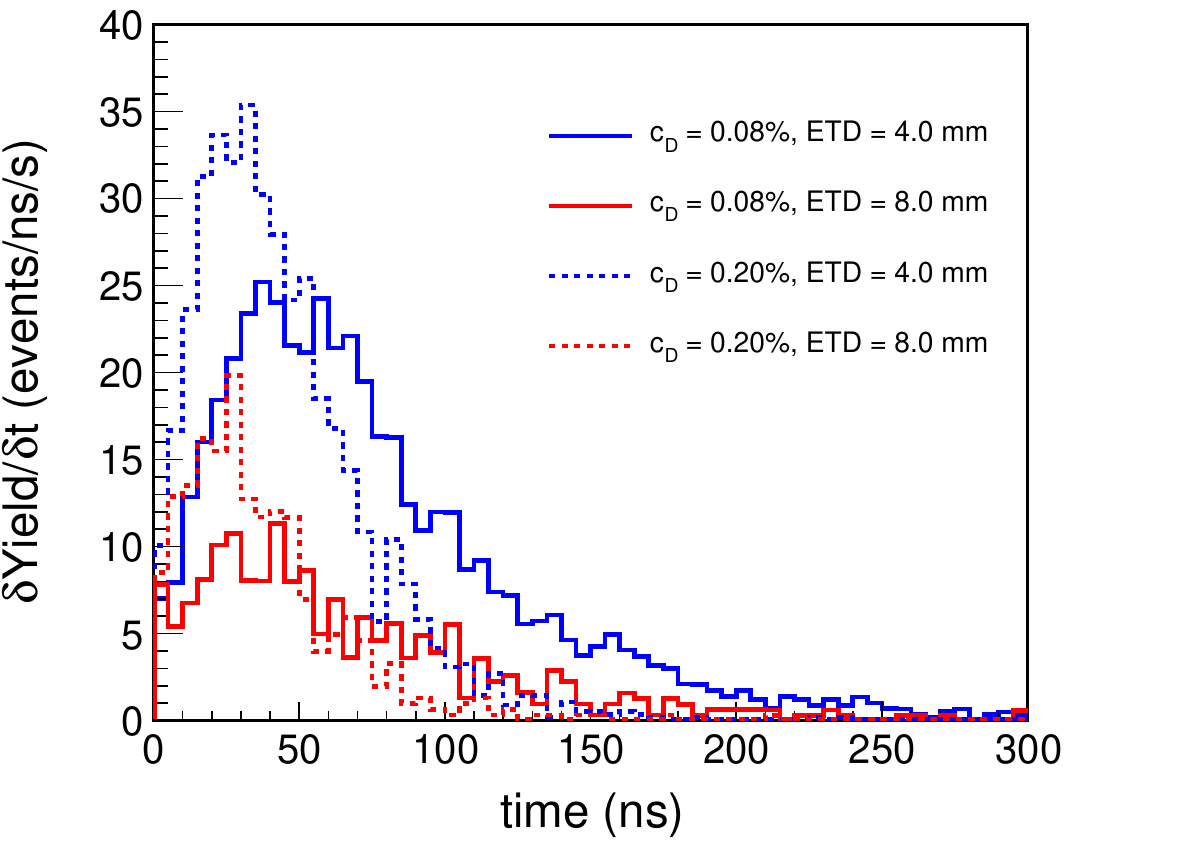}
    \caption{The time distribution of muonic atoms of hydrogen reaching the target for two $c_{\text{D}}$ values and for two $ETD$ values. }
    \label{fig:time-distr}
\end{figure}

\begin{figure*}
    \centering
    \includegraphics[width=1\linewidth]{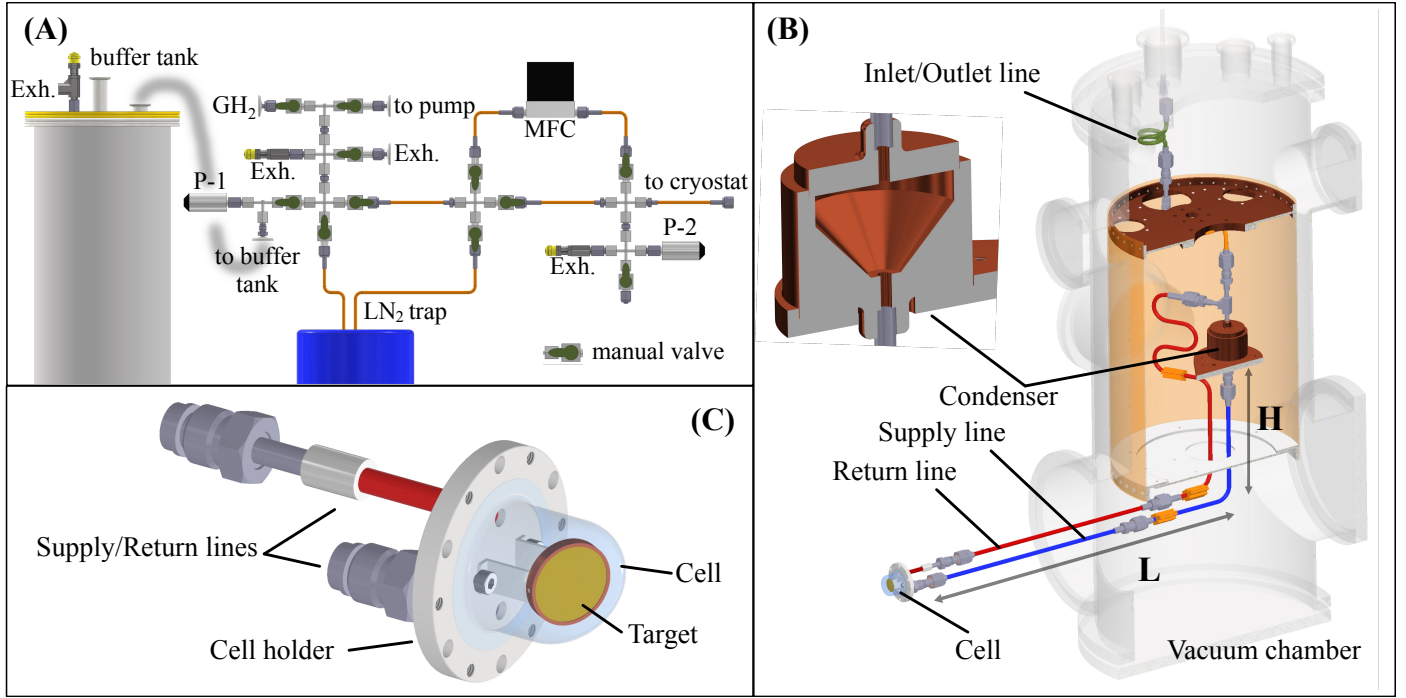}
    \caption{(A) The schematic of the gas handling system. The buffer tank, the GH$_2$ source inlet, the LN$_2$ trap, the mass flow controller (MFC), manual valves, safety valves leading to exhaust (Exh.) and pressure sensors (P-1 and P-2) are labeled. (B) The drawing of the cryogenic system including the condenser, the Inlet/Outlet, Supply, and Return pipes, and the cell. The height (H) and the horizontal length (L) of the cryogenic arm are labeled. The full cryogenic assembly is placed inside a vacuum chamber with an additional copper radiation shield around the second stage of the cold head and the condenser. (C) The 3D drawing of the liquid hydrogen cell with the target for the muonic x-ray spectroscopy placed inside. }
    \label{fig:cryogenic-sys}
\end{figure*}
As a measure of efficiency and transfer rate we refer to the following two quantities: efficiency and yield. The efficiency is computed as the number of muonic deuterium atoms reaching the target divided by the number of muons passing through the entrance scintillator. The yield is defined as the number of muonic deuterium atoms reaching the target per second for a realistic muon beam profile accessible at PSI; the muon beam intensity as a function of the muon momentum follows the official reported values by PSI and is scaled for an average primary proton beam of 2~mA (the maximum reported primary proton beam intensity is 2.4~mA) for a realistic absolute transfer rate evaluation.

The relation between the muon momentum and the ETD parameter can be seen in Figure~\ref{fig:EDT-muP}. Moreover, we can observe that the efficiency has its maximum for low ETD and it goes as high as 16\% (with DEST = 25~mm). The maximum efficiency value drops to 13\% with DEST = 65~mm and down to 10\% with DEST = 105~mm. Figure~\ref{fig:eff-Yield_cD} shows the efficiency (left) and the yield (right) for selected pairs of optimum ETD and muon momentum, as a function of the deuterium concentration on the x-axis. In these plots, it becomes visible that although the efficiency is the largest for the low ETD values, the beam intensity dependence on the muon momentum at PSI compensates such that the yield is slightly lower for low ETD and then saturates for larger ETD values, remaining constant. As a results, the optimum ETD value for muon transfer is around 4~mm, this is where the yield increases high enough and going to larger ETD does not bring any benefit; on the opposite, higher ETD and higher muon momentum mean higher beam intensity and implicitly larger background level due to muon decays. For the ETD of 4~mm, the efficiency becomes 10.1~\% and the yield is 669.6~s$^-1$ (with DEST = 25~mm); for DEST increased to 105~mm, the efficiency drops to 7.5~\%. From the same two plots we can select the optimum deuterium concentration, in this case $\sim$0.1\% (0.06-0.12\%).

Compared to the optimum deuterium concentration value for the hydrogen gas cell reported in Ref.~\cite{Adamczak2023} of 0.25\%, the present optimum deuterium concentration for hydrogen in liquid state is lower. This is a direct consequence of the higher density of liquid hydrogen ($0.071 \text{ g/cm}^3$) compared to the 100-bar gas ($0.008 \text{ g/cm}^3$). Because of this higher overall density, a lower relative deuterium fraction achieves the absolute number density required for the initial muon transfer, while simultaneously minimizing $\mu d - d$ scattering in the highly collisional liquid medium to prevent kinetic energy loss and preserve the enhanced diffusion range driven by the Ramsauer-Townsend effect.\\

The time distribution of the muons reaching the target relative to the moment when they pass through the entrance scintillator is plotted in Figure~\ref{fig:time-distr}. For the time distribution corresponding to the parameters $c_{\text{D}}=0.08\%$ and $ETD=4$mm (with its corresponding optimized muon momentum value), the mean time is 75~ns, the time corresponding to the peak of the distribution is 37.5~ns, and 90\% of muon transfers to the target happen within 149~ns after the muon triggers the entrance scintillator. This amounts to considerable reduction (more than fourfold) in the timescale compared to the hydrogen gas cell~\cite{Adamczak2023}. The muon transfer timescale can further be reduced by using a larger deuterium concentration, see case of $c_{D}=0.20\%, ETD=4mm$ in Figure~\ref{fig:time-distr}. In this case the efficiency is slightly reduced (9.4~\%), but the time distribution will have a mean of 44~ns, the time corresponding to the peak of the distribution is 32.5~ns, and 90\% of muon transfers to the target happen within 83~ns after the muon triggers the entrance scintillator. The reduction in the time scale of the muon diffusion and transfer process translates into the possibility to impose narrower time conditions for obtaining the coincidence x-ray spectrum triggered by the muons detected in the entrance scintillator, which increases the signal-to-background ratio.
\begin{figure*}
    \centering
    \includegraphics[width=1\linewidth]{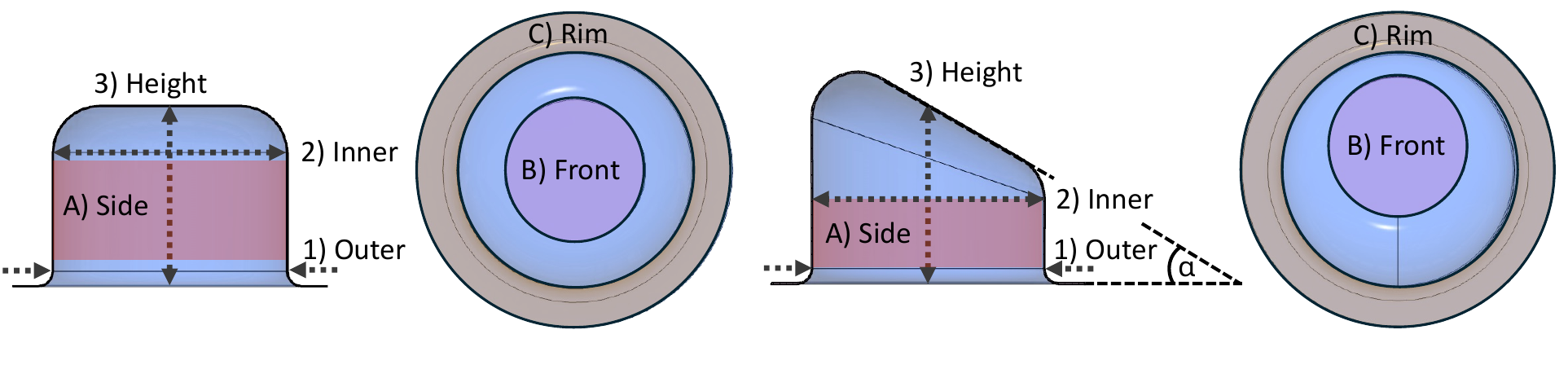}
    \caption{Schematics for the visualization of positions for the measured dimensions of the flat (left) and tilted (right) cells.}
    \label{fig:cell-dimensions}
\end{figure*}
\begin{table*}
    \centering
    \caption{Measured dimensions of cells for the positions shown in Figure~\ref{fig:cell-dimensions}. The mean values are accompanied by the standard deviation given in round parentheses obtained over 20 flat cells and 10 tilted cells, each with 10 measuring points around the cell for each listed dimension.}
    \label{tab:cell-dimensions}
    \begin{tabular}{ccc | ccc}
        Dimension & Flat (mm) & Tilted (mm) & Dimension & Flat ($\mu$m) & Tilted ($\mu$m)\\
        \hline
         Outer Diameter (1) & 30.77(5) & 30.90(4) & Side Thickness (A) & 157(7) & 174(6)\\
         Inner Diameter (2) & 30.08(14) & 30.10(8) & Front Thickness (B) & 175(5) & 178(5) \\
         Height (3) & 23.64(23) & 23.37(51) & Rim Thickness (C) & 174(6) & 177(5) \\
         \hline
         Volume (ml) & 16.66(16) & 16.51(36) & Angle $\alpha$ (°)& - & 27.90(94)\\
         \hline
    \end{tabular}
\end{table*}

\subsection{Conclusions on muon transfer in liquid hydrogen}
To conclude the muon transfer simulation and the comparison of liquid hydrogen and gaseous hydrogen cells: we find similar muon transfer efficiency in the optimum geometrical configuration between the two types, liquid-based and gas-based hydrogen cells. The optimum deuterium concentration decreases to 0.1\% in this case compared to 0.25\% for the gas hydrogen cell. An interesting feature is the time distribution of muonic deuterium atoms reaching the target for transfer: the time scale of the muon diffusion and transfer in liquid hydrogen is found to be shorter compared to gaseous hydrogen, allowing for a better background rejection for the prompt x-ray spectrum. And finally the most important aspect remains the capability to operate the liquid hydrogen cell at atmospheric pressure compared to 100~bar for gaseous hydrogen as in Ref~\cite{Adamczak2023}, while the densities in the two cases are $0.071 \text{ g/cm}^3$ and $0.008 \text{ g/cm}^3$, respectively. This allows us to use thin-walled cells with liquid hydrogen, drastically lowering the attenuation of low-energy muonic x-rays, of interest for low-mass atoms. For example, for a 30~keV x-ray, the attenuation would be 5.1\% with the liquid hydrogen cell and its design for the enclosure (out of which 3.7\% attenuation comes from an average length of LH$_{2}$ of 15~mm crossed by x rays emitted from the target, 0.724\% from the Mylar cell enclosure, and an additional 0.724\% attenuation from the vacuum chamber Mylar window), while for the current design of the hydrogen gas cell~\cite{Adamczak2023}, the attenuation is closer to 50\% (which does not affect high-energy x rays from the high-mass elements of focus for muX). Aside from attenuation, the line shape is also highly affected if the material budget between the target and the detector is not kept low, by inducing Compton scattering in this material and creating a low-energy tail in the observed x-ray peaks.

\section{Technical description}
\subsection{Cryogenic system}
The cryogenic system and the cell are designed as the scaled-down version of the liquid hydrogen target of the STRASSE project used for proton-induced nuclear reactions~\cite{Liu2023,Enciu2026}. The liquid hydrogen is circulated to the cell on the principle of the thermosiphon loop as in Ref.~\cite{Enciu2026}. For liquefying the hydrogen, a two-stage cold head is used (model PT415 Cryomech) which can deliver 15-20~W power at the second stage in the temperature range of 15-22~K and reaches a base temperature of 3.3~K at the second stage. At the second stage, a condenser vessel is mounted in which hydrogen liquefies and acts as the heat sink of the thermosyphon loop, as shown in Figure~\ref{fig:cryogenic-sys}. Two heaters of LH-HTR-25-100 type from Lakeshore are connected in series and used for keeping the temperature of the condenser at the setpoint temperature chosen by the user, in the 15-22~K range, controlled by a Lakeshore 335 temperature controller module. Silicon diode temperature sensors (DT-670) are used for monitoring the temperature throughout the cryogenic system (first stage of the cold head, second stage of the cold head, supply line, return line, two sensors on the stainless steel holder of the cell -- top and bottom), while PT-103 sensors are used on the warmer/passive components such as the radiation shield and as checks for the thermally-decoupled mechanical support structures. The hydrogen gas is supplied to the cryogenic system from a gas handling system shown in Figure~\ref{fig:cryogenic-sys}. A buffer tank of 40~L is used to store the hydrogen gas. A nitrogen trap is placed between the buffer tank and the pipe filling the cryogenic system for discarding impurities such as CO$_2$, H$_2$O, and C$_n$H$_m$ present in the hydrogen gas as contaminants at levels of $<$0.5~ppm. A mass flow controller, placed between the nitrogen trap and the line filling the cryostat, keeps track of the total amount of hydrogen filled into the system. Additionally, pressure transducers and safety relief valves are mounted to form a complete monitoring and safety system.

\subsection{Liquid hydrogen transfer cell}
\label{sec:Cell}
The cell holding the liquid hydrogen is made out of thermoformed Mylar (biaxially oriented polyethylene terephthalate -- BoPET). The selection of Mylar for the transfer cell material provides several distinct advantages over conventional metal target vessels. First, Mylar is composed of low atomic number ($Z$) elements which suppresses background muonic x-rays that could otherwise obscure spectral lines of interest from low-$Z$ implanted targets. Second, the high mechanical tensile strength of biaxially oriented PET at cryogenic temperatures enables the fabrication of thin-walled geometries (down to $\approx 100~\mu\text{m}$) capable of safely withstanding moderate pressure at cryogenic temperature. Most importantly, these thin polymeric windows exhibit negligible x-ray absorption in the soft energy regime ($< 50\text{ keV}$), which is critical for high-resolution spectroscopy on light to medium-$Z$ muonic atoms with Metallic Magnetic Calorimeters (MMCs). The cell body is thermoformed into a single monolithic shell, thereby eliminating lateral seams, reducing risk of cryogenic vacuum leaks, and presenting a uniform thickness profile across the beam and detection axes. See Figure~\ref{fig:cryogenic-sys} for an overview of the design and geometry of the cell.

For the production of the cell, the initial thickness of the Mylar foil, before thermoforming, is 179(2)~$\mu$m. A flat circular film of Mylar is warmed-up up to 200$^{\circ}$C and then pressed in the desired shape using a mold holding the foil and a press-head with the geometry of the final cell shape. The cell is made out of a single Mylar foil. After cell production, it is glued on the stainless steel holder using an epoxy adhesive (3M DP190 gray) as in Ref.~\cite{Enciu2026}.

Two types of cells are produced, one with a flat front face and one with a tilted front face. The final dimensions of the cells are shown in Figure~\ref{fig:cell-dimensions} and Table~\ref{tab:cell-dimensions}. For the flat cells, the average outer diameters and inner diameters were \SI{30.77 +- 0.05}{\milli\meter} and \SI{30.08 +- 0.14}{\milli\meter}, respectively, compared to \SI{30.90 +- 0.04}{\milli\meter} and \SI{30.10 +- 0.08}{\milli\meter} for tilted cells. The diameters were measured using a caliper. After measuring the cell at 4 different height positions, the larger than expected difference between the outer and inner diameter was attributed to a linear slope in the cell diameter along its height. The flat cell height was also measured directly using the caliper, yielding \SI{23.70 +- 0.15}{\milli\meter}, in agreement with the value of \SI{23.64 +- 0.23}{\milli\meter} obtained by volume measurement. Due to the geometry of the tilted cells, a direct measurement of the height was not possible; instead, their height was determined indirectly from the water filling volume. Finally, a thickness characterization of the cell was performed with a micrometer, yielding a thinner side wall thickness, likely due to stretching during thermoforming.

The cells are tested for robustness under several conditions: burst test at room temperature, pressure test at cryogenic temperature, and a low-pressure check for each cell before operation. For each produced batch of cells, flat and tilted, up to ten cells are filled with water at room temperature until they burst. The bursting pressure turned out to be 24-26~bar for both types, with no systematic difference between the two different cell shapes. At cryogenic temperatures burst tests performed with liquid nitrogen (77~K) lead to a minimum bursting pressure of 3.8~bar absolute pressure. After each assembly of the setup for a new operation of the liquid hydrogen setup, pressurized air around 1-2~bar overpressure is injected in the cell, while the cell is in the chamber under vacuum. The stability of the inside pressure and the vacuum level in the chamber are monitored for ensuring tightness and robustness of the cell before operation with liquid hydrogen.

\begin{figure}[t]
    \centering
    \includegraphics[width=1\linewidth]{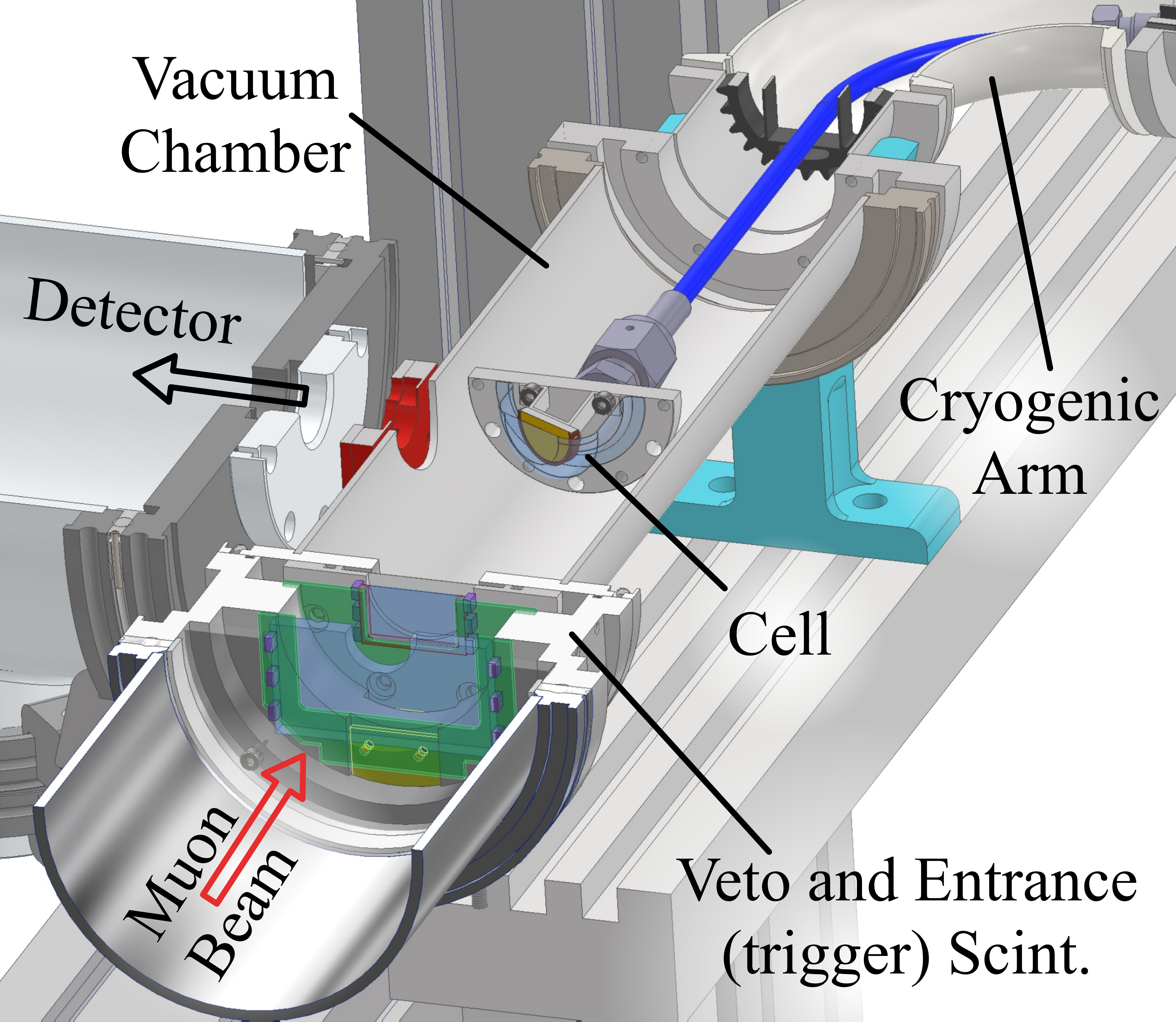}
    \caption{Section view of the liquid hydrogen cell system integrated on the beamline and coupled with the metallic magnetic calorimeter cryogenic detector. This assembly configuration is used for muonic x-ray spectroscopy using the transfer method on low-mass atomic nuclei. The cryogenic arm is bent due to space constraints, the cryostat being located to the right of the beamline axis. The distance between the target and the entrance (trigger) scintillator detector is 105~mm. Several mechanical support components, fasteners, and thermal insulation components are not shown for clarity of the image.}
    \label{fig:QUARTET-integration_closeup}
\end{figure}
\begin{figure*}[t]
    \centering
    \includegraphics[width=1\linewidth]{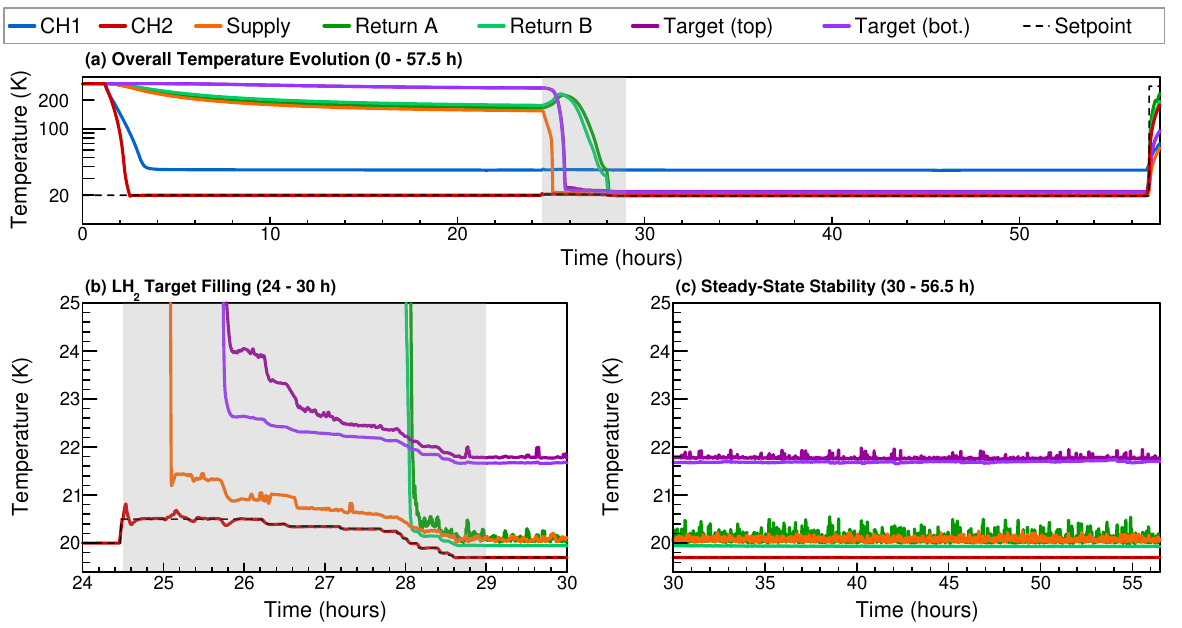}
    \caption{Measured cooldown curves showing the temperature evolution of the cold head stages, condenser, supply line, return lines, and target cell holder over time (a). The gray region shows the the time interval between the moment of hydrogen filling into the cryostat and the moment the cell became full with liquid hydrogen. The warm-up procedure is initiated after the 56.5 hours of operation. Plots (b) and (c) show a zoom-in for the phases of target filling and of steady-state stable operation.}
    \label{fig:temp-cryo-plots}
\end{figure*}
\begin{figure*}[t]
    \centering
    \includegraphics[width=0.475\linewidth]{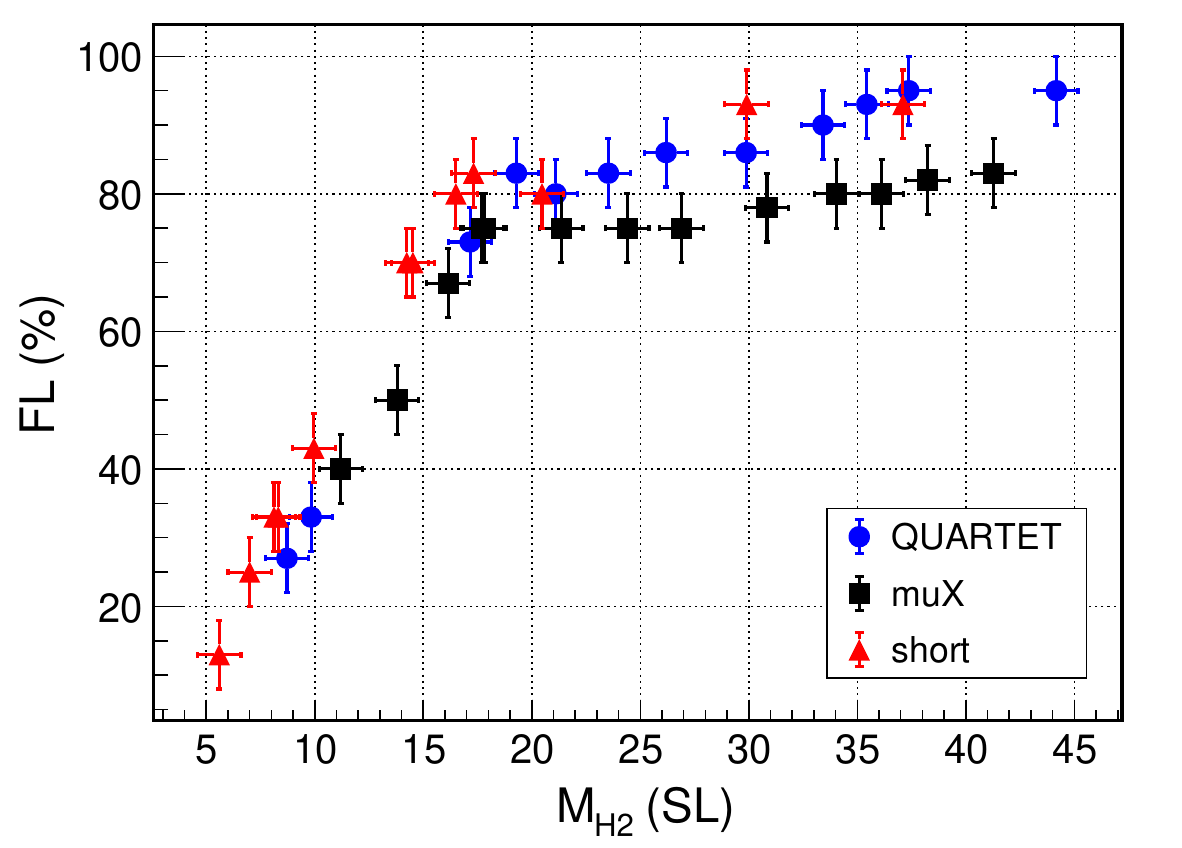}
    \includegraphics[width=0.475\linewidth]{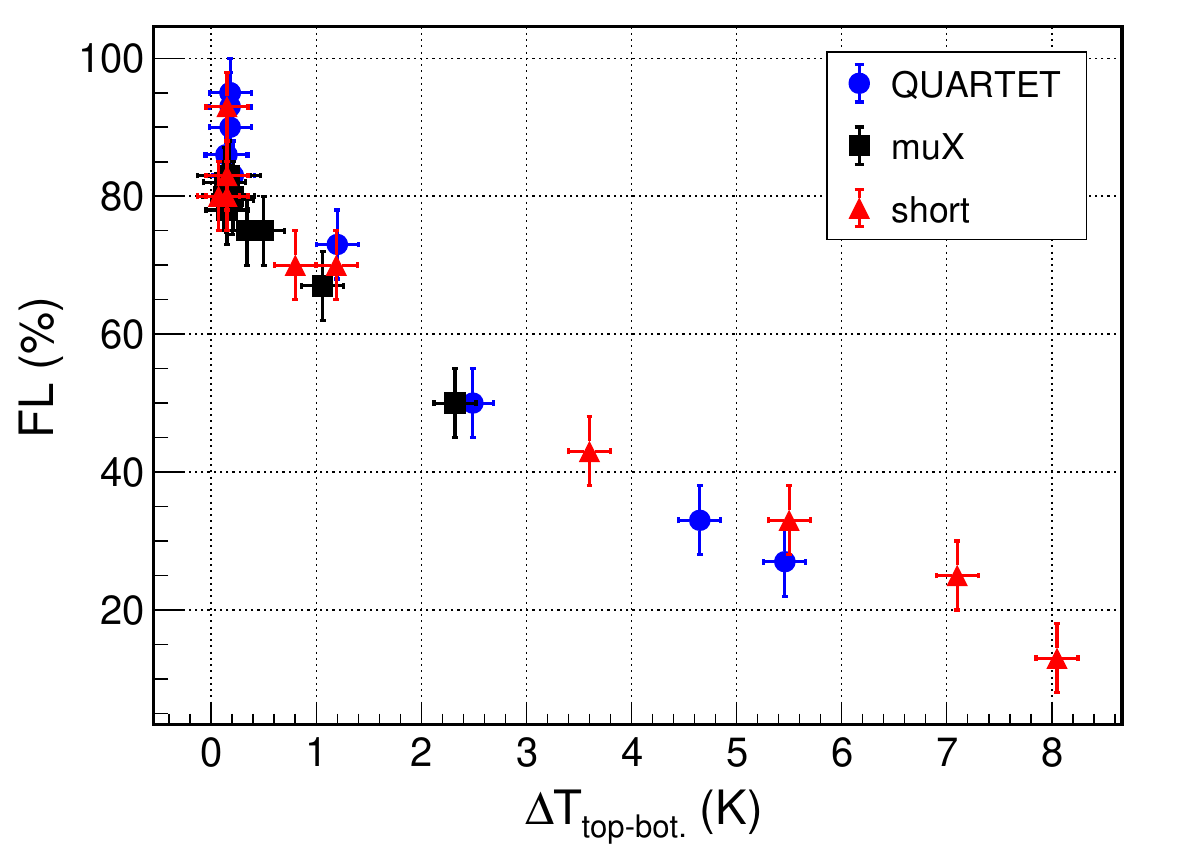}
    \caption{Left: The relation between the filling level of liquid hydrogen in the cell (FL) and the amount of hydrogen introduced in the cryostat (M$_{\text{H2}}$). Right: The relation between FL and the temperature difference recorded on the top and the bottom position on the cell holder ($\Delta T_{\text{top-bot.}}$). The error bars correspond to the maximum uncertainty evaluated for the FL, M$_{\text{H2}}$, and $\Delta T_{\text{top-bot.}}$, of 5\%, 1~SL, and 0.2~K, respectively.}
    \label{fig:cryo-plots}
\end{figure*}
\section{System integration within the full experimental setup}
\label{system-integration}
This section presents the integration of the liquid hydrogen system within the full experimental setup for muonic x-ray spectroscopy. The cell full of hydrogen, containing the target used for the spectroscopy measurement, needs to be situated in front of the muon beam and surrounded by the x-ray detection system. This typically implies that the cryostat of the liquid hydrogen setup must be situated far away from the hydrogen cell due to mechanical constraints.

Figure~\ref{fig:QUARTET-integration_closeup} shows how our hydrogen cell is placed in the experiments when an MMC~\cite{Unger2024}, accompanied by its own refrigerator system, as in the QUARTET experiments~\cite{Ohayon2024}. The cell with a tilted shape is used and placed in front of the muon beam, with the tilted face of the cell oriented towards the MMC. The two cryostat systems are situated on either side of the beamline, with the supply and return pipes of the liquid hydrogen system being bent at 90$^{\circ}$. The cryogenic arm in this geometrical configuration reaches a length of 703~mm with a chosen curvature radius of 70~mm. The distance between the cell and the condenser is 522~mm along the perpendicular direction to the beamline, 262~mm along the beamline direction, and a height difference of 218~mm, or 230~mm relative to the supply pipe level.

The liquid hydrogen target cell system is primarily designed for QUARTET experiments, but it can also be adapted for use with an HPGe array as in muX experiments~\cite{Wauters2021,Adamczak2023}. In this case the cryogenic arm presents no bending, but it needs a greater length in order for the cell to be situated in the center of an HPGe array such as the GIANT~\cite{Gerchow2023}, ISOLDE Decay Station (IDS)~\cite{Fynbo2017}, or similar HPGe array systems amounting to $\sim$2.0~m total diameter considering their respective supporting frames. In this case, the length of the cryogenic arm extends up to 1.3~m, keeping the same height difference between the cell centerline and the condenser bottom of 218~mm.

From the thermal performance point of view, the insulation becomes the most critical part along the cryogenic arm extension in order to keep the hydrogen integrally liquid until it reaches the cell. Insulation composed of alternating high-reflectivity aluminum foil and vacuum-safe insulator fabric layers was used around the supply line and return line individually (4-8 layers each) and additional $\sim$ 4 layers around both supply and return lines. The pipes were guided to remain centered inside the vacuum tubing using 3D-printed (PLA/PETG) centering pieces with a spike-geometry to minimize the contact to the vacuum tubing and the pipes carrying liquid hydrogen. The cell was fixed at its position inside the vacuum chamber also using 3D-printed components (PLA/PETG). With these insulation considerations, the cell was successfully filled with liquid hydrogen in all geometrical configurations described in this section.


\section{Commissioning and performance}
The cryogenic setup was tested in three different lengths and geometries of the cryogenic arm. The first configuration has a short cryogenic arm as in Figure~\ref{fig:cryogenic-sys}, with a horizontal length of pipes of only 450~mm. The second configuration of the cryogenic arm is shown in Figure~\ref{fig:QUARTET-integration_closeup}, compatible with the MMC detector with an arm length of 703~mm, and the third configuration is with a 1.3-m long straight cryogenic arm compatible with an HPGe array as described in Section~\ref{system-integration}.

In all cases, the cryostat reaches the end-point temperature at the second stage and at the first stage of 20~K and 37~K, respectively, within 2~h of cryostat operation. After a total of 10~h of cryostat operation, the supply pipe, return pipe, and the cell holder reach 170~K, 180-190~K, and 270~K, respectively; these components will not cool efficiently any further unless hydrogen is filled in as a medium of heat exchange. With hydrogen gas filled inside the cryogenic system, it will immediately liquefy inside the condenser and will reach the cell via the supply line within 1~h after H$_2$ filling, with a full cell achieved after another four hours of cooling. The time evolution plot of the temperature at the different measuring points along the thermosiphon loop is shown in Figure~\ref{fig:temp-cryo-plots}. 

The filling level of the liquid hydrogen in the cell is controlled by the amount of hydrogen filled in the cryogenic system (measured in standard liters, SL, equivalent volume at standard temperature and pressure of 1 atm and 20$^{\circ}$C), as shown in Figure~\ref{fig:cryo-plots}~(left) for the three different geometrical configuration of the cryogenic arm. The filling level is determined by visual inspection using a webcam attached to the vacuum chamber window; it is defined as the percentage of the height of the liquid level relative to the total diameter of the cell and it is extracted from digital photographs.

The supply line is positioned on the bottom of the cell holder and the return line is positioned on the top of the cell holder. The holes on the cell holder, where the supply and return pipes are welded on, are distributed 24~mm apart from each other. The return line continues to be routed at the same height until it bends upwards below the condenser. The height of the return pipe relative to the bottom edge of the cell is 27~mm considering the pipe center or 25~mm considering the pipe lower edge. This height within the cell is equivalent to a filling level of $FL=90\%$ (27~mm/30~mm) or $FL=83.3\%$ (25~mm/30~mm), respectively. This causes the upper 5~mm of the cell, for $FL>83.3\%$ to contain bubbles and little liquid as the fluid needs to fill the entire horizontal return pipe at that height level. Figure~\ref{fig:cryo-plots} shows exactly this: the filling level increases steadily with the increase in the amount of hydrogen inserted in the cryogenic system until $FL\approx83\%$ where the increase of $FL$ changes slope. Beyond $FL\approx83\%$, the cell does not manage to fill up as quickly when more hydrogen is introduced in the system because the full return line needs to first be filled. The configuration of $muX$ is the most affected one, as seen in Figure~\ref{fig:cryo-plots}, because the cryogenic arm is the longest in this case. From the design point of view, an $S$-shape return pipe, as done in Ref.~\cite{Enciu2026}, would be ideal for a 100~\% filling level. Nevertheless, it was not implemented due to space constrains as it would imply a cryogenic arm of larger diameter. Moreover, the complete filling of the cell well above the target region does not affect the muon transfer process in the current geometry of the cell.

To be noted is that bubbles can also form along the supply line, and in that case they have to escape towards the return line by passing through the cell. For this reason, bubbles can be visible in the cell, coming from the inlet and rising upwards to the outlet, at the back of the cell's body and behind the target mounted in the cell. As the relevant region for the muon transfer is the layer of hydrogen between the entrance of the muons in the cell and the target, the bubbles present in the back of the cell do not affect the muon transfer. There, any fine bubble would only result in a negligible net lowering of the density of hydrogen by less than $0.01\%$.

An observable correlated with the filling level which can be used in experimental conditions, when the cell cannot be visually monitored, is the difference between the temperature reading of the two sensors mounted on the cell holder. One sensor ($top$) is mounted near the return line on the back of the cell holder and one sensor ($bot.$) is mounted near the supply line on the back of the cell holder. The filling level as a function of the $\Delta T_{\text{top-bot.}}$ difference in temperature is plotted in Figure~\ref{fig:cryo-plots}~(right). One can observe a perfect agreement for the dependence of the filling level on $\Delta T_{\text{top-bot.}}$ between the present setup and the one in Ref.~\cite{Enciu2026}. Despite the two systems being different on several aspects (different cell holder geometries and masses, total H$_2$ quantities, and cell sizes), they share the same cell diameter size (30.08~mm in the present work and 31~mm in Ref.~\cite{Enciu2026}) and the same vertical distance between the supply and return line on the cell holder welding point ($\approx$24~mm).

\section{Conclusions}
In this work, we have presented the design, simulation, and cryogenic commissioning of a novel liquid hydrogen target transfer cell tailored for precision muonic x-ray spectroscopy on microgram-quantity targets. By operating with liquid hydrogen at atmospheric pressure rather than high-pressure gaseous hydrogen for increased density, the setup enables the use of thin ($175~\mu\text{m}$) thermoformed Mylar cell walls. This configuration drastically reduces low-energy x-ray attenuation, limiting 30 keV photon absorption down to 5\% and opening new experimental territory for spectroscopy on light nuclei ($Z \le 12$) using Metallic Magnetic Calorimeter (MMC) detector arrays. Monte Carlo simulations demonstrate that the high density of liquid hydrogen ($0.071\text{ g/cm}^3$) achieves optimal transfer efficiencies exceeding 10\% (and total yields up to $670\text{ s}^{-1}$ at PSI beam intensities) at an optimal cell layer thickness ($\text{ETD} \approx 4\text{ mm}$) and an admixture of $c_D \approx 0.1\%$. Furthermore, the higher atomic density shortens the muonic diffusion timescale with 90\% of transfer events occurring within less than 150~ns, a value that can be reduced down to less than 85~ns for $c_D \approx 0.2\%$. This rapid diffusion time window will allow exceptionally narrow coincidence gating relative to the beam trigger scintillators, providing a considerable reduction in accidental background. Cryogenic testing across three separate arm geometries confirmed reliable thermosiphon operation. The liquid filling level of the cell correlates directly with the vertical temperature difference ($\Delta T_{\text{top-bot}}$) across the cell holder, providing a robust, non-visual diagnostic of liquid target coverage during beamline operation. After a complete in-beam commissioning, the setup will be positioned for upcoming experimental campaigns at the Paul Scherrer Institute, enabling benchmark-quality measurements of absolute charge radii of rare or radioactive targets.

\section*{Acknowledgments}
M.~E. was funded by an ETH Postdoctoral Fellowship during this work. This research is supported by the NCCR Muoniverse, a National Centre of Competence in Research, funded by the Swiss National Science Foundation (grant number 51NF-0\_229254). The authors acknowledge the support of the mechanical workshops of IPA, ETH Z\"urich and Paul Scherrer Institute for building this experimental setup.\\

\bibliographystyle{unsrtnat}
\bibliography{references}

\end{document}